\documentclass{ws-ijmpe}
\usepackage[compress]{cite}
\usepackage[colorlinks=true,linktocpage=true,linkcolor=blue,citecolor=blue,allcolors=blue]{hyperref}
\usepackage[utf8]{inputenc}

\newcommand{\eq}[1]{\begin{align} #1 \end{align}}

\begin{document}

\newcommand{\titul}{Hadron resonance gas with van der Waals interactions}

\markboth{V. Vovchenko}{\titul}

\catchline{}{}{}{}{}

\title{\titul}

\author{Volodymyr Vovchenko\footnote{\emph{Present address}: Nuclear Science Division, Lawrence Berkeley National Laboratory, 1 Cyclotron Road, Berkeley, CA 94720, USA}}

\address{Institut f\"ur Theoretische Physik,
Goethe Universit\"at Frankfurt,\\ D-60438 Frankfurt am Main, Germany}

\address{Frankfurt Institute for Advanced Studies, Giersch Science Center,\\ D-60438 Frankfurt am Main, Germany}

\maketitle

\begin{history}
\end{history}

\begin{abstract}
An overview of a hadron resonance gas (HRG) model that includes van der Waals interactions between hadrons is presented. 
Applications of the excluded volume HRG model to heavy-ion collision data and lattice QCD equation of state are discussed.
A recently developed quantum van der Waals hadron resonance gas model is covered as well. 
Applications of this model in the context of the QCD critical point are elaborated.
\end{abstract}

\keywords{hadron resonance gas; excluded volume; van der Waals interactions.}



\section{Introduction}

\begin{figure}[th]
\centerline{\includegraphics[width=.6\textwidth]{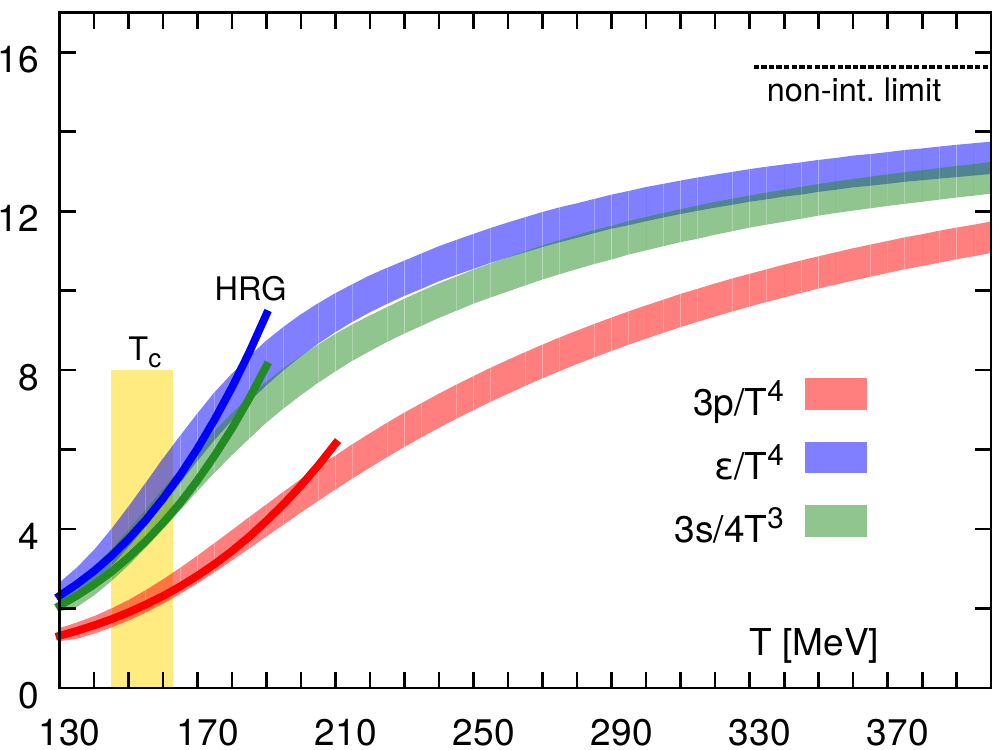}}
\caption{From Ref.~\cite{Bazavov:2014pvz}.
Scaled pressure~(red band), energy density~(blue band), and entropy densiy~(green band), evaluated in lattice QCD at the physical point by the HotQCD collaboration.
The solid curves correspond to the ideal HRG model predictions, the dashed line depicts the Stefan-Boltzmann limit of massless quarks and gluons, the yellow band corresponds to the chiral pseudocritical temperature.
}
\label{fig:lqcdEos}
\end{figure}

The equation of state of strongly interacting matter is in the focus of investigations performed in heavy-ion collision experiments, and in lattice gauge theory~[see e.g.~\cite{Busza:2018rrf,Bzdak:2019pkr,Ratti:2018ksb} for recent reviews].
First-principle lattice QCD simulations suggest a smooth crossover transition at vanishing net baryon density~\cite{Aoki:2006we}, from a dilute hadron gas type matter at low temperatures to a phase which has thermodynamic properties similar to those of a quark-gluon plasma~[see Fig.~\ref{fig:lqcdEos}].
The transition is characterized by the chiral pseudocritical temperature $T_{pc} \simeq 155$~MeV~\cite{Borsanyi:2010bp,Bazavov:2011nk}, below which one expects to see a confined hadron phase.
This phase is usually described by the hadron resonance gas (HRG) model.
This model, its roots going back to Hagedorn~\cite{Hagedorn:1965st}, essentially assumes that hadronic interactions are dominated by the resonance formation.
In the simplest and most commonly used model variant -- the ideal HRG model -- the system is modeled as a non-interacting, multi-component gas of known hadrons and resonances.
Arguments based on the S-matrix formulation of statistical mechanics do suggest that the inclusion of resonances as additional free particles describes those attractive hadronic interactions which result in the formation of narrow resonances~\cite{Dashen:1969ep}.
This picture has also been supported by the phase shift analysis of certain hadron-hadron scattering data~\cite{Venugopalan:1992hy}.

The ideal HRG model describes quite well both, the lattice QCD observables at temperatures below $T_{pc}$~\cite{Borsanyi:2011sw,Bazavov:2012jq,Bellwied:2015lba,Bellwied:2013cta}, as well as the hadron multiplicities in relativistic heavy-ion collisions at various energies~[see e.g. \cite{Andronic:2017pug} for a recent review].
At the same time, hadronic interactions exist that are not dominated by resonance formation.
Repulsive interactions are often modeled through an excluded volume~(EV) correction of van der Waals~(vdW) type, employing the notion of hadron eigenvolumes introduced long time ago in the context of the Hagedorn bag-like models~\cite{Baacke:1976jv,Hagedorn:1980kb}.
HRG models with repulsive interactions have recently received renewed interest in the context of lattice QCD data on fluctuations and correlations of conserved charges~\cite{Albright:2015uua,Huovinen:2017ogf,Vovchenko:2017xad}. 
In particular, it was pointed out that deviations of the lattice data on higher-order susceptibilities from the uncorrelated hadron gas baseline can be attributed to repulsive interactions.
Recent applications of the EV-HRG model are covered in this review.

\begin{figure}[th]
\centerline{\includegraphics[width=.6\textwidth]{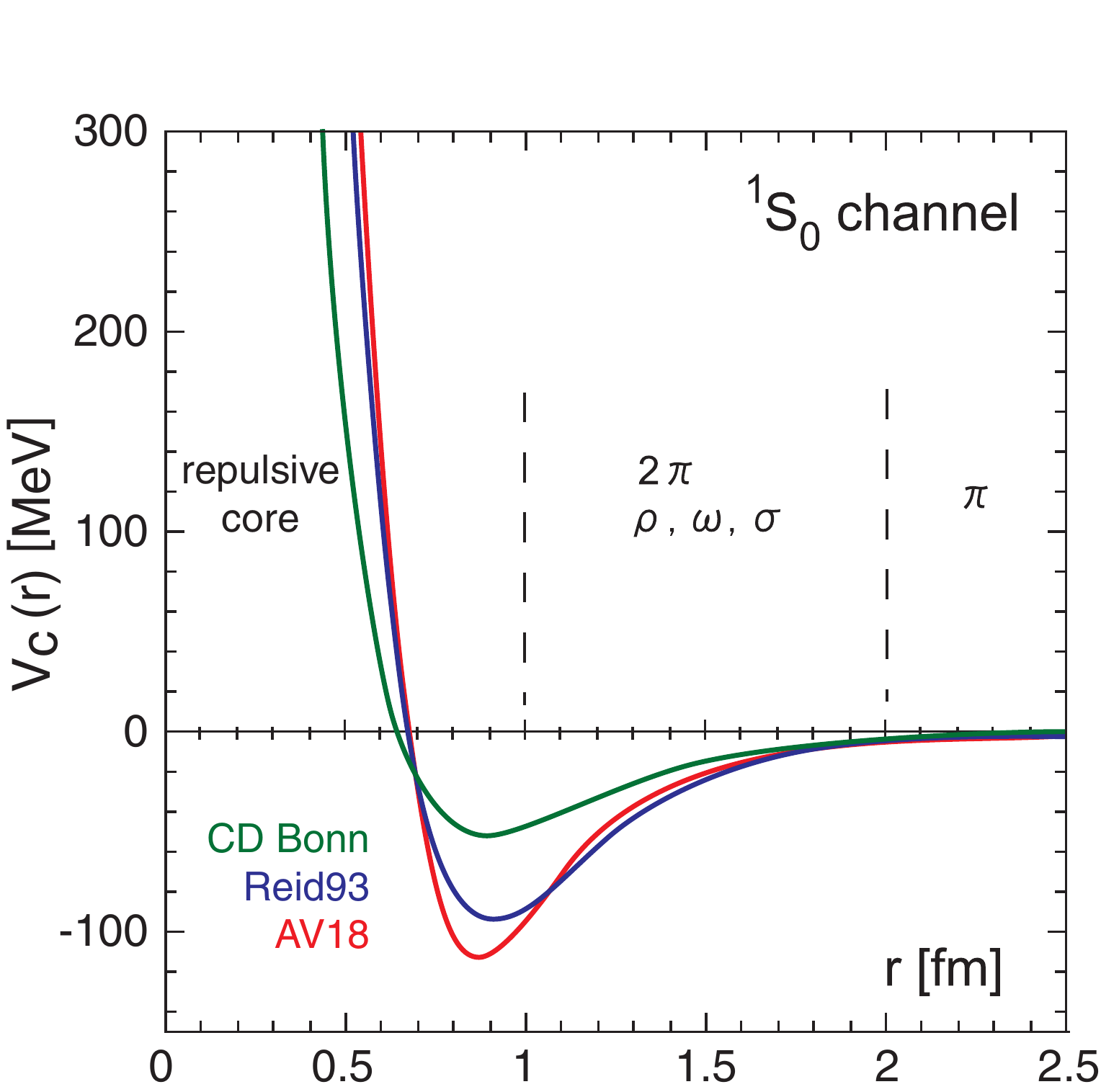}}
\caption{From Ref.~\cite{Ishii:2006ec}.
The nucleon-nucleon interaction potential in the $^1S_0$ channel, evaluated in three models constrained to nucleon-nucleon scattering data: CD~Bonn~(green line)~\cite{Machleidt:2000ge}, Reid93~(blue line)~\cite{Stoks:1994wp}, 
and AV18~(red line)~\cite{Wiringa:1994wb}.
}
\label{fig:NNpot}
\end{figure}

In addition to repulsive interactions, some of the attractive interactions cannot be described by simply adding resonances as free particles.
One example is the nucleon-nucleon interaction at an intermediate range, which is commonly attributed to scalar meson exchange.
The phenomenological nucleon-nucleon potential, shown in Fig.~\ref{fig:NNpot}, exhibits a hard-core type repulsion at short range, and an attractive well at intermediate range -- the structure of the interaction potential common for many atomic and molecular systems.
The vdW equation of state is a simple model which incorporates both of the above features.
Recently, a quantum van der Waals equation was developed, which additionally incorporates effects of quantum statistics and provides a reasonable description of basic nuclear matter properties~\cite{Vovchenko:2015vxa}.
The model was then generalized to incorporate vdW interactions in a full HRG~\cite{Vovchenko:2016rkn}, opening new applications such as studying the effects of criticality in the QCD phase diagram.
We cover here recent developments on that avenue.

\section{Ideal HRG model}

In the simplest setup, the conjectured hadronic phase is described by a multi-component,
ideal gas of point-like hadrons -- the ideal HRG model.
In the grand canonical ensemble (GCE) formulation of the ideal HRG 
the system reduces to an \emph{uncorrelated gas of hadrons}.
Thus, the pressure is given by
\eq{\label{eq:PHRGid}
p^{\rm id}_{\rm hrg}(T,\mu) = \sum_i p_i^{\rm id}(T,\mu_i) =
\lim_{V \to \infty}
\frac{T}{V} \sum_i \ln Z_i^{\rm id} (T,V,\mu),
}
where the sum goes over all hadron species included in the model,
$Z_i^{\rm id} (T,V,\mu)$ and
$p^{\rm id}_i (T, \mu_i)$ are the grand partition function and pressure of the ideal 
Fermi or Bose 
gas at the corresponding temperature and chemical potential for species $i$.
$p^{\rm id}_i (T, \mu_i)$ reads
\eq{\label{eq:piid}
p^{\rm id}_i (T, \mu_i) = \frac{d_i}{6\pi^2} \int_0^{\infty} \frac{k^4 dk}{\sqrt{k^2+m_i^2}}
\left[ \exp\left(\frac{\sqrt{k^2+m_i^2} - \mu_i}{T}\right)+\eta_i\right]^{-1}~.
}
Here $d_i$ and $m_i$ are, respectively, the spin degeneracy factor and mass of hadron species $i$, and $\eta_i$ determines the statistics, being equal +1 for fermions, -1 for bosons, and 0 for Boltzmann approximation.

Other thermodynamic quantities are given by expressions similar to~\eqref{eq:PHRGid}, namely, as a sum over the corresponding ideal gas quantities for all hadron species. 
The particle density of hadron species $i$ is $n_i^{\rm id}(T,\mu_i)$, i.e. it is simply given by
the ideal gas relation for species $i$:
\eq{\label{eq:nid}
n^{\rm id}_i (T, \mu_i) = \frac{d_i}{2\pi^2} \int_0^{\infty} k^2 dk
\left[ \exp\left(\frac{\sqrt{k^2+m_i^2} - \mu_i}{T}\right)+\eta_i\right]^{-1}~.
}
The Boltzmann approximation~($\eta_i = 0$) is sufficient for many HRG applications. In this case Eq.~\eqref{eq:nid} simplifies to
\eq{
n^{\rm id}_i (T, \mu_i) = d_i \, \phi(T, m_i) \, e^{\mu_i/T}, 
\qquad
\phi(T, m) = \frac{m^2\,T}{2\pi^2}\, K_2(m/T).
}
Here $K_2$ is the modified Bessel function of the 2nd order.

Only the light unflavored and strange hadrons are considered here. 
Within the GCE formulation, all conserved charges, such as
the baryon number $B$, the electric charge $Q$, and the net strangeness $S$,
are conserved on average.
Therefore, there are three corresponding independent chemical potentials: $\mu_B$, $\mu_S$, and $\mu_Q$. The chemical potential of the $i$th hadron species is thus determined as 
\begin{equation}
\mu_i\ =\ B_i\,\mu_B\, +\, Q_i\,\mu_Q\, +\, S_i\,\mu_S \, ,
\label{eq:mui}
\end{equation}
with $B_i = 0,\, \pm 1$, $Q_i = 0,\, \pm 1,\, \pm 2$, and
$S_i = 0,\, \pm 1,\, \pm 2,\, \pm 3$ being the corresponding conserved charges of the $i$th hadron species: baryon number, strangeness, and electric charge. 
The notation $\mu$ will be used to denote all chemical potentials, $\mu\equiv (\mu_B,\mu_S,\mu_Q)$.

The finite widths of the resonances are often incorporated into the model,
through an additional integration of the thermal functions over the mass distribution $f_i(m)$~\cite{Becattini:1995if,Wheaton:2004qb,Torrieri:2004zz}:
\eq{
n_i^{\rm id}(T,\mu_i) \Rightarrow \int d m\, f_i(m)\, n_i^{\rm id}(T,\mu_i).
}
The mass distribution $f_i(m)$ is commonly taken in a Breit-Wigner form with an energy-(in)dependent width.
A recent comparable study suggests that details of the finite widths modeling notably influence the densities of broad resonances, which can be important in precision thermal model applications~\cite{Vovchenko:2018fmh}.
Other recent developments include modeling the resonances through the derivatives of the empirical hadron scattering phase shifts~\cite{Huovinen:2016xxq,Fernandez-Ramirez:2018vzu}, or through the K matrix formalism~\cite{Wiranata:2013oaa,Dash:2018can}.
As the resonance formation is not directly linked to the van der Waals interactions, we omit further discussion of the resonance widths in this review.

\begin{figure}[t]
\centering
\includegraphics[width=.49\textwidth]{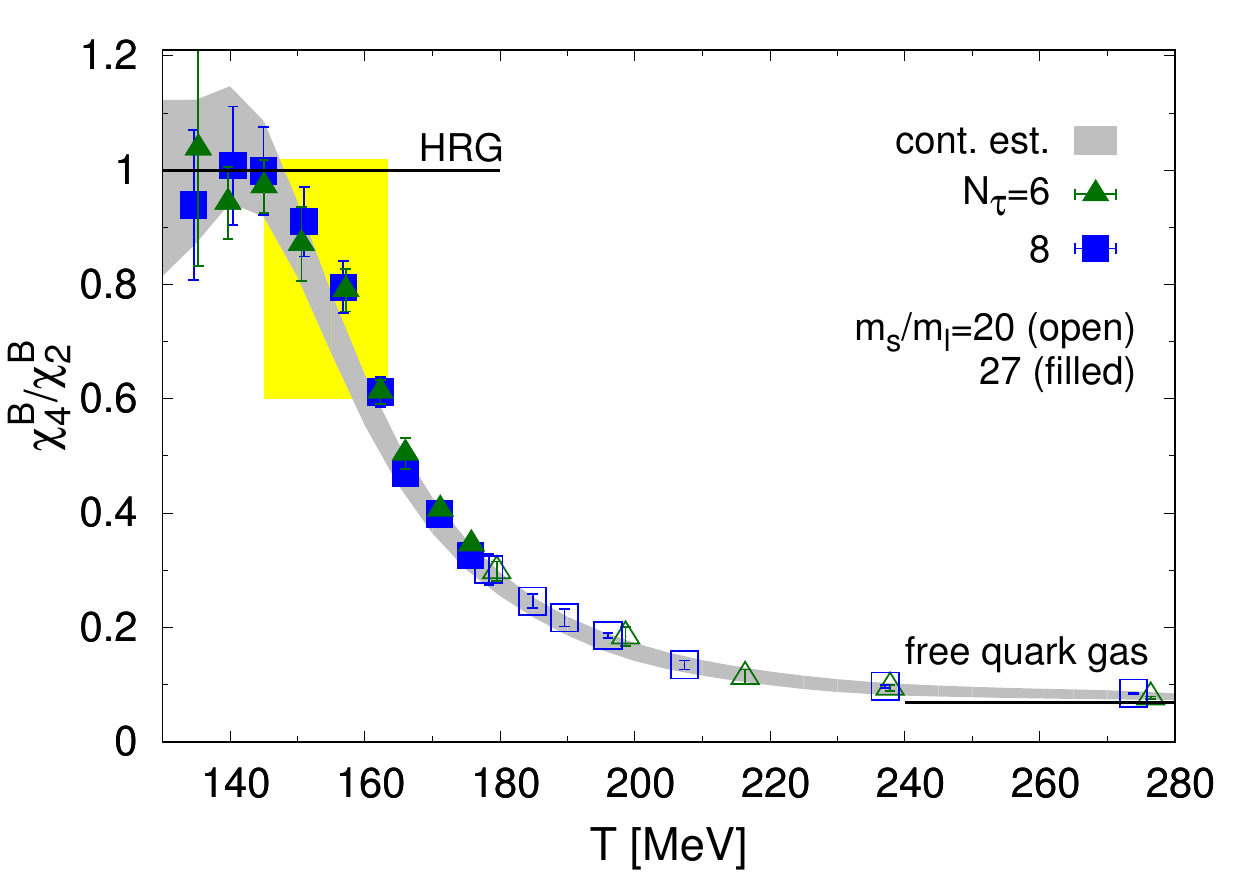}
\includegraphics[width=.49\textwidth]{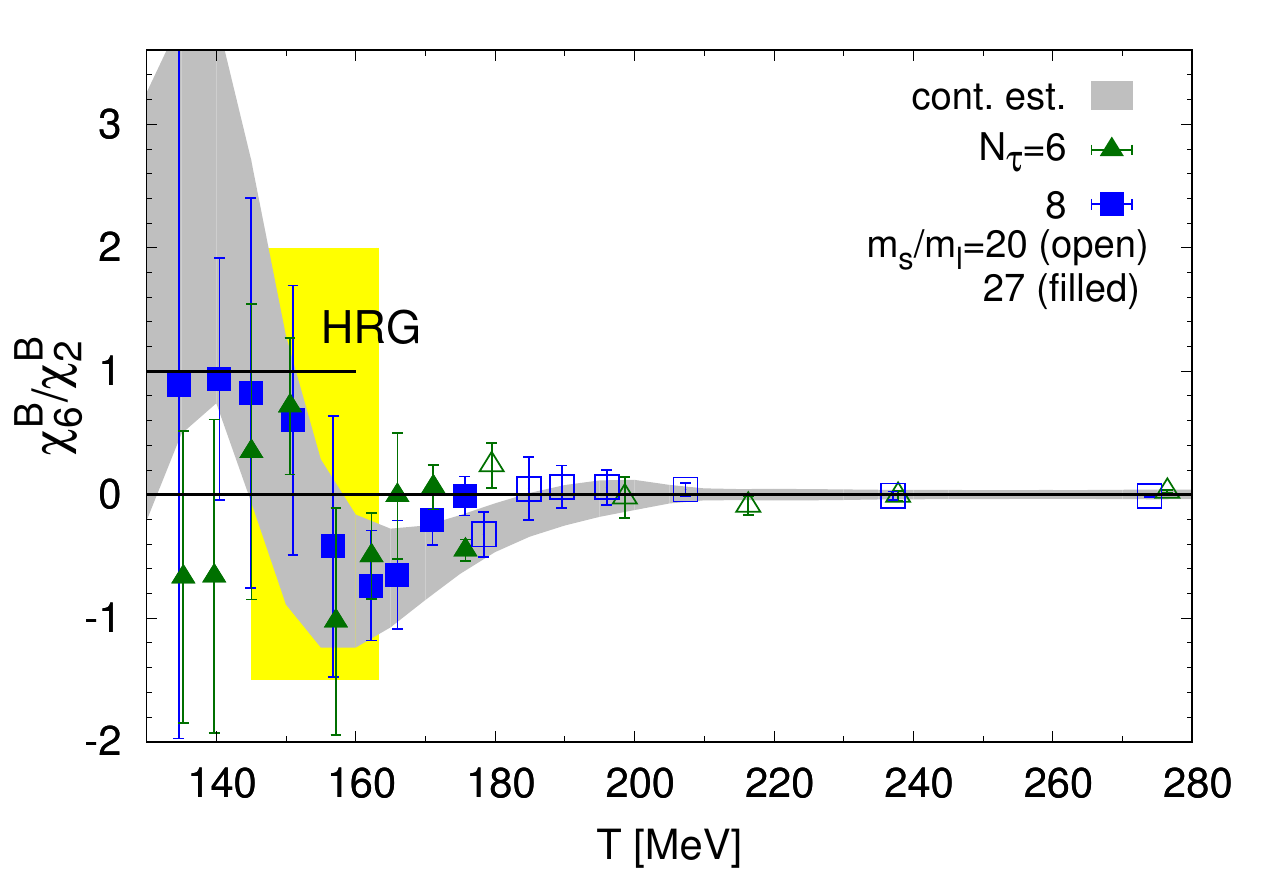}
\caption{From Ref.~\cite{Bazavov:2017dus}.
The temperature dependence of the $\chi_4^B/\chi_2^B$~(\emph{left panel}) and $\chi_6^B/\chi_2^B$~(\emph{right panel}) ratios, evaluated in lattice QCD by the HotQCD collaboration.
The horizontal solid lines correspond to the ideal HRG model baseline.
}
\label{fig:chi4chi6}
\end{figure}

While the ideal HRG model describes  the thermodynamic functions of QCD at $\mu_B = 0$ well at temperatures up to~(and even somewhat above) $T_{pc}$, the situation is different for susceptibilities. These are defined as derivatives of the pressure function with respect to the chemical potentials,
\eq{\label{eq:susc}
\chi_{lmn}^{BSQ}~=~\frac{\partial^{l+m+n}p/T^4}{\partial(\mu_B/T)^l \,\partial(\mu_S/T)^m \,\partial(\mu_Q/T)^n}~\,,
}
and thus probe the finer details of the equation of state.
In recent years, the susceptibilities of conserved charges have been computed in lattice QCD at the physical point~\cite{Borsanyi:2011sw,Bazavov:2012jq}.
They show rapid deviations from the uncorrelated has of hadrons picture in the vicinity~(and even below) $T_{pc}$.
A particularly transparent example is given by the ratios of certain susceptibilities.
The baryon number kurtosis ratio, $\chi_4^B / \chi_2^B$, is equal to unity in an uncorrelated gas of hadrons\footnote{Up to small corrections due to Fermi statistics.}, irrespective of the input hadron spectrum or resonance widths modeling.
This is due to the fact that no hadron with a multiple baryon number is present in an HRG\footnote{As long as light nuclei are not considered.}, and in this case the baryon number follows the Skellam distribution -- a difference of two independent Poisson distributed numbers. 
In the case of the ideal HRG these are the numbers of baryons and antibaryons.
The comparison of the ideal HRG baseline for the ratios $\chi_4^B/\chi_2^B$ and $\chi_6^B/\chi_2^B$ with the lattice QCD data is shown in Fig.~\ref{fig:chi4chi6}.
The breakdown of the uncorrelated gas of hadrons picture in the vicinity of $T_{pc}$ is evident.
This further motivates to consider extensions of the ideal HRG model.

\section{HRG with excluded volumes}

One widely used extension of the ideal HRG model is an inclusion of excluded volume corrections.
The concept of hadronic eigenvolumes was first introduced in the framework of the thermodynamic properties of systems with
an exponential mass spectrum, such as the statistical bootstrap model~\cite{Hagedorn:1980kb}, 
or the quark-gluon bag model~\cite{Baacke:1976jv}.
The introduction of finite eigenvolumes for heavy Hagedorn states
allows to remove the Hagedorn limiting temperature phenomenon
and even to obtain a crossover or a first-order phase transition to a quark-gluon plasma~\cite{Gorenstein:1981fa}, which later was shown to be generally compatible with lattice QCD thermodynamics~\cite{Ferroni:2008ej,Vovchenko:2018eod}.
On the lower end of the mass spectrum, the existence of a hard-core appears to be well established for the nucleon-nucleon interaction~(see Fig.~\ref{fig:NNpot}), while evidence for a hard core in nucleon-nucleon, hyperon-nucleon, and hyperon-hyperon interactions is indicated by lattice QCD simulations~\cite{Aoki:2012tk}.

Among the different variants of the EV-HRG model used one can distinguish three groups:

\subsection{EV-HRG model with one common eigenvolume parameter}

This is the simplest variant of the EV-HRG model.
All particles are assumed to have the same size, quantified by a common eigenvolume parameter $v$. 
The effect is modeled through a substitution of the system volume by the available volume, $V \to V - vN$, in the partition function.
The pressure is given by a transcendental equation~\cite{Rischke:1991ke},
\eq{\label{eq:pevhrg}
p(T,\mu) =  \sum_i p_i^{\rm id} (T, \mu_i - v p),
}
while the energy density, entropy density, and all particle number densities are suppressed by a common factor\footnote{Up to a few percent correction due to quantum statistics.}. A derivation of the EV model based on Maxwell's identifies which avoids the use of transcendental equations was presented in Ref.~\cite{Zalewski:2015yea}.

Applications of this model have been considered in the literature in the context of the chemical freeze-out in heavy-ion collisions~\cite{BraunMunzinger:1999qy,Cleymans:2005xv}, the hadronic equation of state~\cite{Andronic:2012ut,Vovchenko:2014pka}, transport coefficients of hadronic matter~\cite{Gorenstein:2007mw,NoronhaHostler:2012ug,Kadam:2015xsa}, and the analytic structure of systems with repulsive interactions~\cite{Taradiy:2019taz}.
The main effect of the EV corrections here is to suppress all the densities relative to the ideal HRG model.
At the same time, the model preserves the ideal HRG model results for ratios of densities of any two species, $(n_i^{\rm ev} / n_j^{\rm ev}) = (n_i^{\rm id} / n_j^{\rm id})$, i.e. the EV effects cancel out in yield ratios.
For this reason the model has no influence on the quality of thermal fits nor on the values of the extracted temperature or chemical potentials from data.
The only effect there is a renormalization of the system volume parameter.

\subsection{EV-HRG with an individual eigenvolume parameter per particle (Diagonal EV-HRG model)}

In a more general Diagonal EV-HRG model~\cite{Yen:1997rv} one can assign a different eigenvolume parameter $v_i$ for each particle species. The pressure is defined by the following transcendental equation:
$$
p(T,\mu) = \sum_i p_i^{\rm id} (T, \mu_i - v_i p),
$$
while the densities are given by
$$
n_i(T,\mu) = \frac{n_i^{\rm id}(T,\mu_i - v_i p)}{1 + \sum_j v_j n_j^{\rm id}(T,\mu_j - v_j p). }
$$

\begin{figure}[t]
\centerline{\includegraphics[width=.6\textwidth]{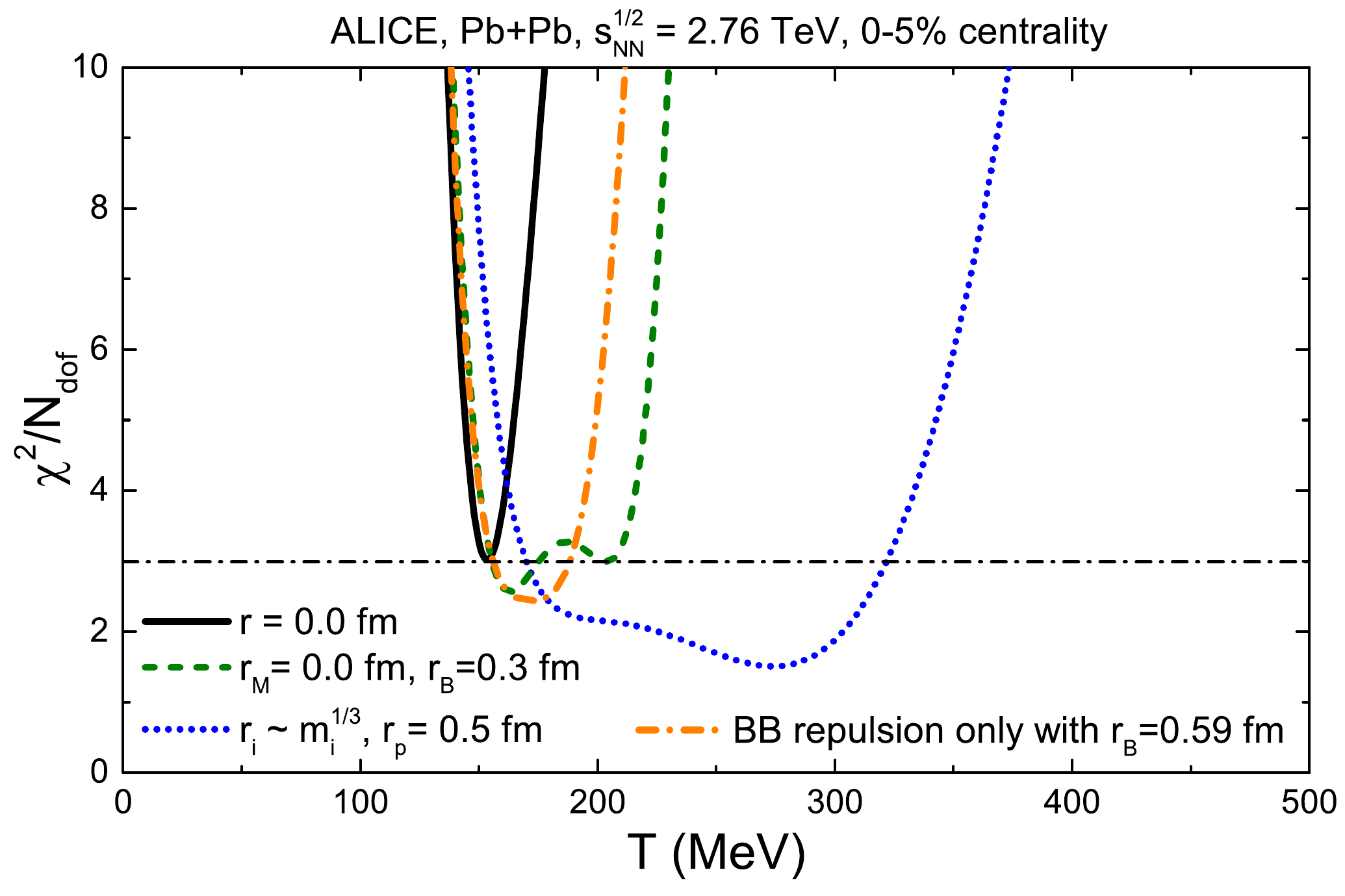}}
\caption{From Ref.~\cite{Vovchenko:2015cbk}.
The $\chi^2$ temperature profiles for thermal fits to 0-5\% ALICE hadron yield data with different variants of HRG model: the ideal HRG model~(black line), the Diagonal EV-HRG model with point-like mesons~(dashed green line) or bag model scaling~(dashed blue line), and the Nondiagonal EV-HRG model with baryon-baryon repulsion only~(dotted blue line).
}
\label{fig:fitsALICE}
\end{figure}

This model is suitable for considering differences in eigenvolumes of different hadrons, e.g. between mesons and baryons. In contrast to the constant eigenvolume scenario, here the yield ratios are modified: in the Boltzmann approximation one obtains $(n_i^{\rm ev} / n_j^{\rm ev}) = (n_i^{\rm id} / n_j^{\rm id}) \, e^{(v_j-v_i)p/T}$, meaning that the EV effects will no longer cancel out in thermal fits to hadron yield ratios in the Diagonal EV-HRG model~\cite{Satarov:2016peb}.
An important implication here is that thermal fits might be affected appreciably, depending on the choice of $v_i$'s.
This was illustrated recently in Ref.~\cite{Vovchenko:2015cbk} for different cases, including the bag model scaling, $v_i \sim m_i$~(radii scale as $r_i \sim m_i^{1/3}$), or point-like mesons~($v_m = 0$), as shown in Fig.~\ref{fig:fitsALICE} for the resulting $\chi^2$ temperature profiles.
One generally observes broadening of the $\chi^2$ minima, which sometimes end up at very high temperatures, $T \gtrsim 200$~MeV.
Description of the QCD matter in terms of hadrons at such high temperatures is certainly questionable and these results should not be viewed as revised values of the chemical freeze-out conditions. 
The correct values of the EV parameters for all the species are not well established either.
Therefore, these results motivate further investigations of the EV effects in a hadron gas.

Another consideration is a possibility of flavor hierarchy in hadron eigenvolumes, in particular differences between strange and non-strange hadrons.
A recent analysis of lattice QCD data on second and higher order susceptibilities of conserved charges prefers smaller eigenvolumes of strange hadrons compared to light flavored ones~\cite{Alba:2017bbr}.

Applications of the Diagonal EV-HRG model to heavy-ion data can also be found in Refs.~\cite{Yen:1997rv,Yen:1998pa,Vovchenko:2015cbk,Alba:2016hwx}.
Other applications of this model include the equation of state~\cite{Steinheimer:2010ib,Andronic:2012ut,Begun:2012rf,Albright:2014gva} and fluctuation observables~\cite{Fu:2013gga,Bhattacharyya:2013oya,Albright:2015uua,Alba:2017bbr}.

\subsection{HRG with an eigenvolume parameter per each pair of particle species (Nondiagonal EV-HRG model)}

An even more general framework to include excluded-volume interactions in a multi-component system is the Nondiagonal EV-HRG (NDEV-HRG) model~\cite{Gorenstein:1999ce,Vovchenko:2016ebv,Satarov:2016peb}, where the repulsive interactions are introduced for each pair of particle species, in a form of the matrix $\tilde{b}_{ij}$ of the excluded volume parameters.
The total pressure is partitioned into the sum of ``partial'' pressures,
\eq{
p(T,\mu) = \sum_i p_i (T,\mu),
}
which are determined by the following system of transcendental equations.
\eq{\label{eq:NDE:pi}
p_i (T,\mu) = p_i^{\rm id} (T, \mu_i^*), \qquad \mu_i^* = \mu_i - \sum_j \widetilde{b}_{ij} \, p_j, \qquad i = 1, \ldots , f,
}
These equations are generally solved numerically.
The particle number densities $n_i \equiv (\partial p / \partial \mu_i)_T$ are found as the solution to the system of linear equations
\eq{\label{eq:NDE:ni}
\sum_j [\delta_{ij} + \tilde{b}_{ji} \, n_i^{\rm id} (T, \mu_i^*)] \, n_j = n_i^{\rm id} (T, \mu_i^*), \qquad i = 1 \ldots f.
}
The entropy and energy densities, as well as other thermodynamic quantities, can be obtained from standard thermodynamic relations.

The Nondiagonal model opens several new applications in hadronic physics.
Most importantly, the model allows to incorporate essential qualitative differences between baryon-baryon, baryon-antibaryon, meson-baryon, and meson-meson interactions.
Whereas the existence of a repulsive core appears to be well established for nucleon-nucleon interactions, this is not necessarily the case for other interactions.
In particular, known baryon-antibaryon interactions at short range are dominated by annihilations rather than repulsions.
The presence of significant mesonic eigenvolumes, comparable to those
of baryons, leads to significant suppression of thermodynamic functions in the crossover region at $\mu_B = 0$, which is at odds with lattice data (see e.g. Refs.~\cite{Andronic:2012ut,Vovchenko:2014pka}).
It can therefore make sense to omit baryon-antibaryon, meson-meson and/or meson-baryon excluded volume interactions while preserving the baryon-baryon ones.
This cannot be achieved within the Diagonal EV model, but is naturally incorporated in the Nondiagonal one, by simply setting the relevant $\tilde{b}_{ij}$ cross terms to zero.
These different possibilities have been studied for a HRG in some detail in Ref.~\cite{Satarov:2016peb}.
The Nondiagonal model reduces to the Diagonal one in a partial case $\widetilde{b}_{ij} \equiv v_i$.
The canonical ensemble formulation of the Nondiagonal EV-HRG model has been studied with the use of Monte Carlo methods~\cite{Vovchenko:2018cnf}.

\subsection{Baryonic eigenvolume and the lattice data}

\begin{figure}[t]
\centering
\includegraphics[width=.49\textwidth]{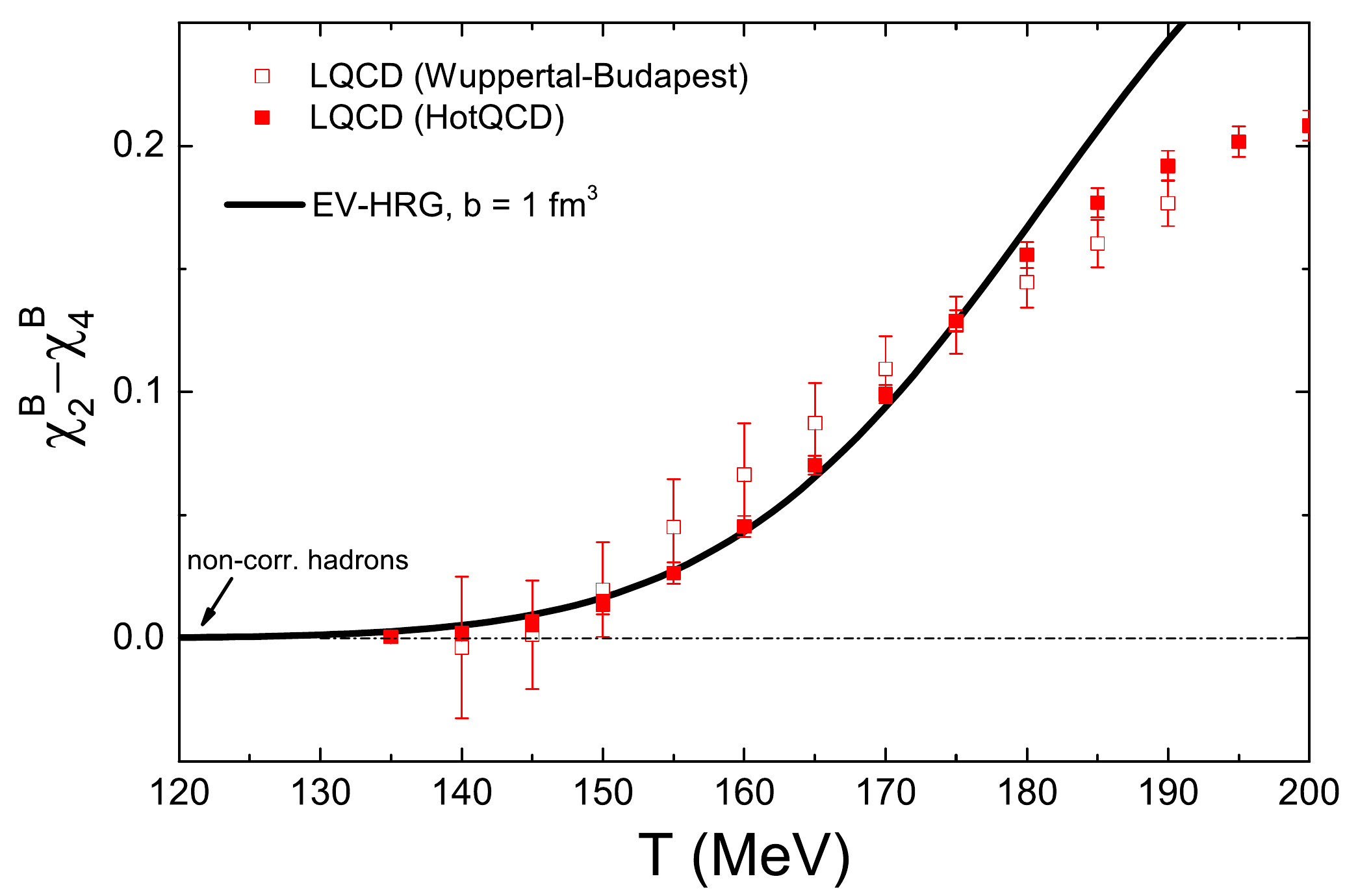}
\includegraphics[width=.49\textwidth]{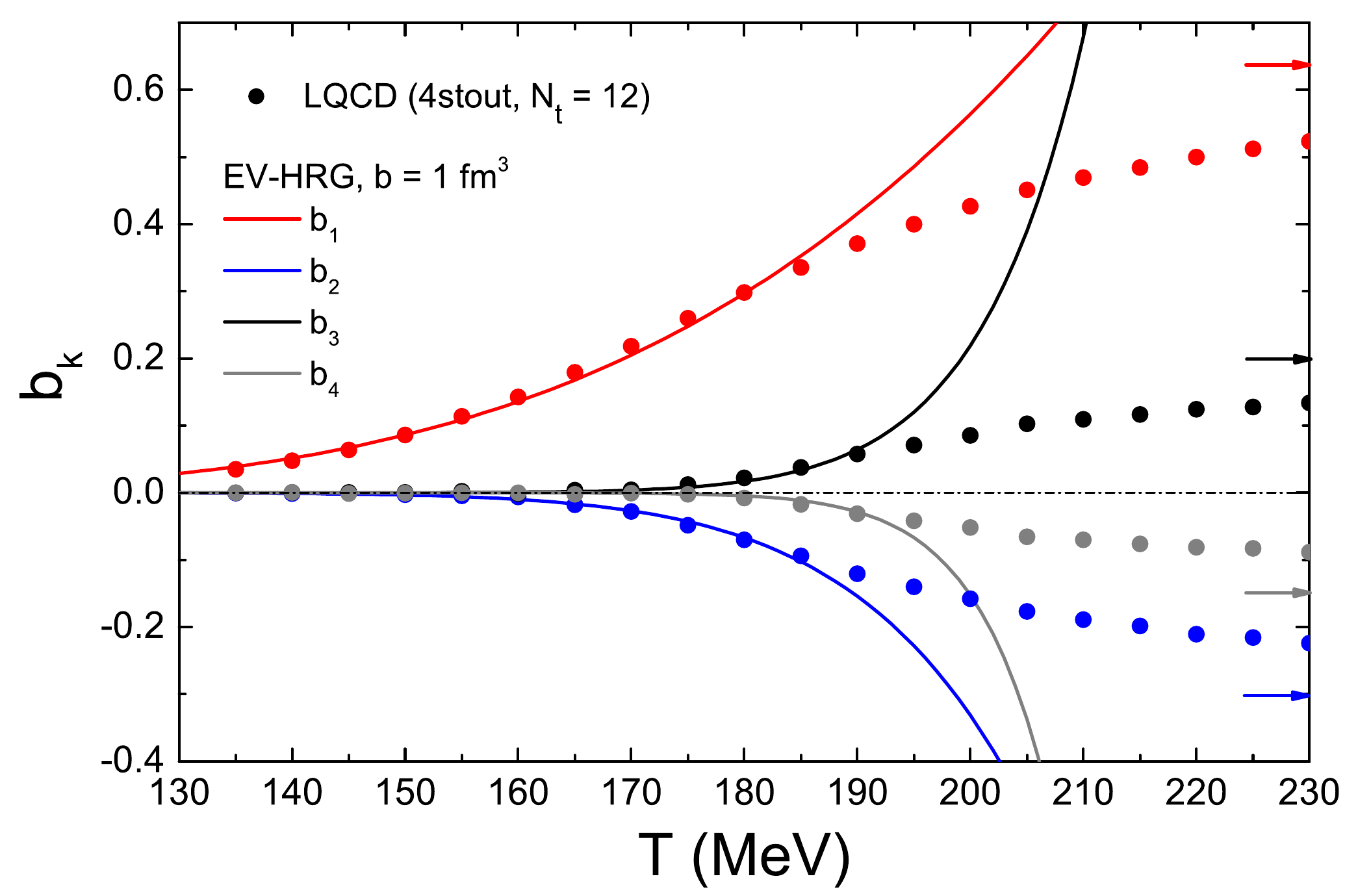}
\caption{
Adapted from Ref.~\cite{Vovchenko:2017xad}.
The temperature dependence of a baryon number susceptibilities difference $\chi_2^B - \chi_4^B$~(\emph{left panel}) and Fourier coefficients $b_1$, $b_2$, $b_3$, and $b_4$~(\emph{right panel}), as evaluated in lattice QCD by the Wuppertal-Budapest~\cite{Bellwied:2015lba,Vovchenko:2017xad} and HotQCD~\cite{Bazavov:2017dus,Bazavov:2017tot} collaborations
and calculated within the EV-HRG model with baryonic eigenvolume parameter $b = 1$~fm$^3$.
}
\label{fig:evhrglqcd}
\end{figure}

The necessity of the EV corrections in a hadron gas has been debated.
Evidence for the existence of baryon-baryon EV-like interactions was obtained from an analysis of recent lattice QCD data on net-baryon susceptibilities $\chi_{2n}^B$~[Eq.~\eqref{eq:susc}] at $\mu_B = 0$ and Fourier coefficients $b_k$ of net-baryon density at imaginary $\mu_B$~\cite{Vovchenko:2017xad}.
For the case of an uncorrelated Maxwell-Boltzmann gas of hadrons one has $\chi_{2n}^B (T) = \chi_2^B (T)$ for all $n\geq 1$ and $b_k(T) = 0$ for $k \geq 2$~(see Ref.~\cite{Vovchenko:2017xad} for details).
The lattice data are consistent with such a behavior at sufficiently low temperatures, $T \lesssim 150$~MeV, whereas deviations set in at higher temperatures. 
The onset of these deviations is captured quite well by a HRG model with baryon-baryon EV interactions, characterized by an eigenvolume parameter $b = 1$~fm$^3$ taken to be common for all species.
This agreement is illustrated in Fig.~\ref{fig:evhrglqcd}.

\subsection{On the connection of the eigenvolume parameter to the hard-core radius}

The EV model is commonly associated with the picture of impenetrable hard spheres, where each particle has a spherical eigenvolume characterized by its hard-core radius.
In classical systems, the eigenvolume parameter $v$ can indeed be connected to the hard-core radius $r$ by calculating the second virial coefficient for a hard-core interaction potential and matching it to the EV model, giving a well-known result $v = (16 \pi/3) \, r^3$~\cite{LL,GNS}.

\begin{figure}[t]
\centerline{\includegraphics[width=.6\textwidth]{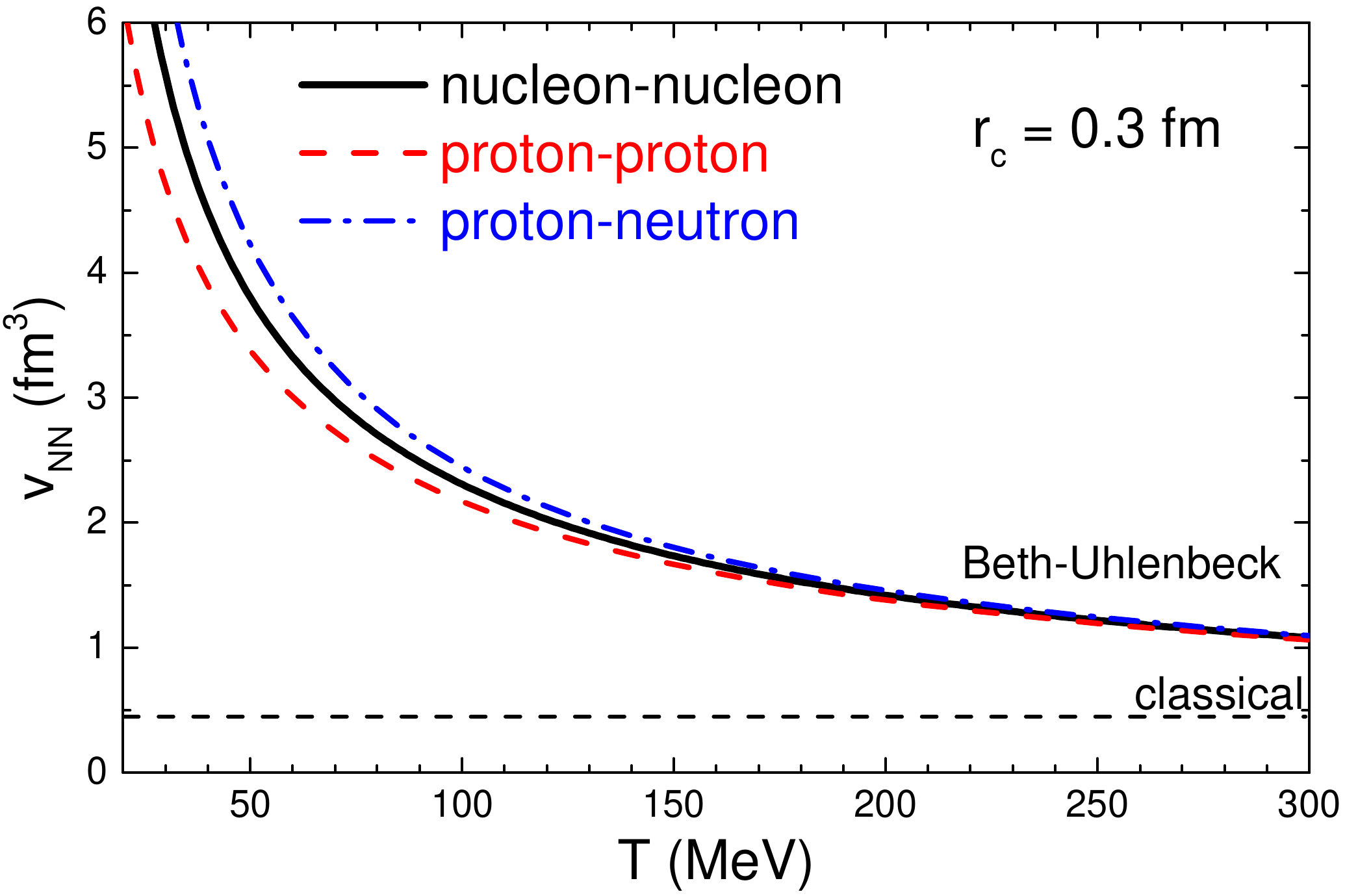}}
\caption{From Ref.~\cite{Vovchenko:2017drx}.
Temperature dependence of the nucleon-nucleon excluded volume parameter $v_{NN}$~(solid
black line), the proton-proton excluded volume parameter $v_{pp}$~(dashed red line), and the proton-neutron excluded volume parameter $v_{pn}$~(dashed red line), as calculated within the
relativistic Beth-Uhlenbeck approach for a hard-core potential with the nucleon hard-core radius of $r_c = 0.3$~fm. 
The dashed horizontal line shows the prediction of the classical hard-spheres model.
}
\label{fig:rhc}
\end{figure}

It should, however, be noted that the above formula connecting the eigenvolume parameter and the interaction hard-core radius is only applicable when quantum mechanical effects can be neglected.
While this is the case for most atomic and molecular systems, the situation is quite different when nuclear and hadronic systems are considered.
It has been suggested, based on Beth-Uhlenbeck approach for non-ideal quantum gases~\cite{Beth:1937zz}, that there are sizable corrections to the classical relation between $v$ and $r$ when hard-core interactions between pions~\cite{Kostyuk:2000nx} or nucleons~\cite{Typel:2016srf} are considered quantum mechanically.
A systematic analysis of this question has been presented in Ref.~\cite{Vovchenko:2017drx} for the case of a system of nucleons interacting via a hard-core interaction potential.
Figure~\ref{fig:rhc} presents temperature dependence of the second virial coefficient characterizing the effect of nucleon-nucleon interactions on the equation of state, evaluated using the quantum mechanical Beth-Uhlenbeck approach~(thick black line)~\cite{Beth:1937zz} and compared to the classical, temperature-independent expression~(horizontal dashed line). 
One can see that the Beth-Uhlenbeck result overshoots
the classical one significantly for all temperatures relevant for hadronic matter. The quantum mechanical calculations approach the classical limit only at unrealistically high temperatures.
This result implies that, because of significant quantum mechanical effects, there is no simple connection between the magnitude of the eigenvolume parameters in the EV-HRG model and the underlying hard-core radii of hadrons.
In that sense, it might be more reasonable to view the EV-HRG model as an effective phenomenological approach to incorporate repulsive interactions into the equation of state, rather than the one based on microscopic properties of hadron-hadron interactions potentials.

\subsection{Light nuclei}

The HRG picture is often extended to incorporate loosely-bound objects such as light (anti-)(hyper-)nuclei~\cite{Andronic:2010qu}.
Within the ideal HRG model these objects are implemented as point particles carrying their quantum numbers and masses. 
The ideal HRG model thus provides essentially a parameter-free description of light nuclei yields at the chemical freeze-out of heavy-ion collisions, which in most cases is in a good agreement with experimental data.
The success of thermal model in describing the light nuclei abundances is quite surprising, as it is rather difficult to imagine that these loosely-bound states with binding energies of few MeV or less can survive in a hot~($T \sim 150$~MeV) medium created in heavy-ion collisions.
An overview of light nuclei production mechanisms can be found in Ref.~\cite{Braun-Munzinger:2018hat}.

The EV-HRG model allows to incorporate the effect of finite sizes of light nuclei by assigning the eigenvolumes to these objects. 
This results in a suppression of their yields relative to the ideal HRG model.
This is similar to the models of nuclear matter equation of state, where the EV correction is often used as a mechanism of cluster dissolution at high densities~\cite{Lattimer:1991nc,Shen:1998by,Typel:2016srf}.
The values of nuclear EV parameters that should be used in a EV-HRG model description are currently not constrained. 
An exploratory study of the EV effects on the production of light nuclei in heavy-ion collisions has been presented in Ref.~\cite{Vovchenko:2016mwg}, indicating a similarly strong sensitivity as in the case of hadrons~(Fig.~\ref{fig:fitsALICE}). 
Analysis of the production of loosely-bound states within a thermal model picture is an active field of research.
An overview of the recent efforts, including the question of EV corrections, can be found in Ref.~\cite{ijmpe_bd}.

\section{HRG with attractive and repulsive van der Waals interactions}

The vdW equation is a simple model describing interacting systems where both the attractive and repulsive interactions are present.
For a single component Maxwell-Boltzmann gas the vdW equation reads
\eq{
p(T,n) = \frac{Tn}{1- b n} - a\, n^2.
}
Here $b$ is the excluded volume parameter characterizing the short-range repulsive interactions. 
This parameter has the same meaning as $v$ in the EV-HRG model~[Eq.~\eqref{eq:pevhrg}].
$a$ is the vdW parameter characterizing intermediate-range attractive interactions through a mean-field approximation. 
A less familiar but a more complete and useful form of the vdW equation is given in terms of the free energy $F$ -- the thermodynamic potential in the canonical ensemble $T,V,N$ variables. The free energy of a van der Waals fluid reads
\eq{\label{eq:Fvdw}
F(T,V,N) = F^{\rm id} (T, V - bN, N) - a \, \frac{N^2}{V}~.
}
Here $F^{\rm id}$ is the free energy of the corresponding fluid in the ideal gas limit.

The most distinctive feature of the vdW equation is the existence of a first-order phase transition with a critical point located at 
\eq{\label{eq:vdwcl}
T_c = \frac{8a}{27b}, \qquad n_c = \frac{1}{3b}, \qquad p_c = \frac{a}{27b^2}~.
}
This fact makes the vdW equation particularly suitable as a toy model for studies of phenomena associated with a critical point.

For QCD applications, several extensions of the classical vdW equation are needed.
First, it is necessary to transform the vdW equation from the canonical ensemble to the grand canonical ensemble, where the number of particles is not fixed. 
This has been achieved in Ref.~\cite{Vovchenko:2015xja}.
Second, effects of quantum statistics need to be accounted for if one wants to apply the vdW equation to describe the ground state of nuclear matter. 
The corresponding quantum van der Waals~(QvdW) equation has been formulated in Ref.~\cite{Vovchenko:2015vxa}~(see also Ref.~\cite{Redlich:2016dpb} for an alternative derivation). The resulting free energy of a QvdW fluid is given by Eq.~\eqref{eq:Fvdw} where $F^{\rm id}$ is the free energy of the corresponding ideal \emph{quantum} gas.
Finally, the model has to also be generalized to describe multi-component systems such as HRG.
The multi-component QvdW equation was developed in Refs.~\cite{Vovchenko:2016rkn,Vovchenko:2017zpj}.

\subsection{Multi-component QvdW model}

The pressure function of the multi-component QvdW model reads
\eq{\label{eq:vdw:p}
p(T,\mu) = \sum_i p^{\rm id}_i (T, \mu_i^*) - \sum_{i,j} \, a_{ij} \, n_i \, n_j.
}
Here the sums run over all species in the particle list.

The particle number densities, $n_i$, satisfy the system of linear equations given by Eq.~\eqref{eq:NDE:ni}
while the shifted chemical potentials, $\mu_i^*$, satisfy the following system of transcendental equations
\eq{\label{eq:vdw:mu*}
\mu^*_i + \sum_j \tilde b_{ij} \, p^*_j - \sum_j (a_{ij} + a_{ji}) \, n_j = \mu_i~, \quad i = 1,\ldots,f~.
}

The pressure, $p(T,\mu)$, at a given temperature $T$ and chemical potentials $\mu$ is determined by first solving numerically the system of equations \eqref{eq:vdw:mu*} for $\mu_i^*$ and then plugging in the result into Eq.~\eqref{eq:vdw:p}.
The entropy and energy densities, as well as various conserved charge susceptibilities, can be obtained using standard thermodynamic relations~[see Ref.~\cite{Vovchenko:2017zpj} for details].

The parameters $\tilde{b}_{ij}$ correspond to the repulsive vdW interactions, and they have the same physical meaning as in the Nondiagonal EV-HRG model.
The parameters $a_{ij}$ correspond to the attractive vdW interactions between hadron species $i$ and $j$, modeled through the mean-field approximation.
The QvdW-HRG model reduces to the Nondiagonal EV-HRG model for the case $a_{ij} \equiv 0$.

\subsection{Nuclear matter as a QvdW system of nucleons}

\begin{figure}[t]
\centering
\includegraphics[width=.49\textwidth]{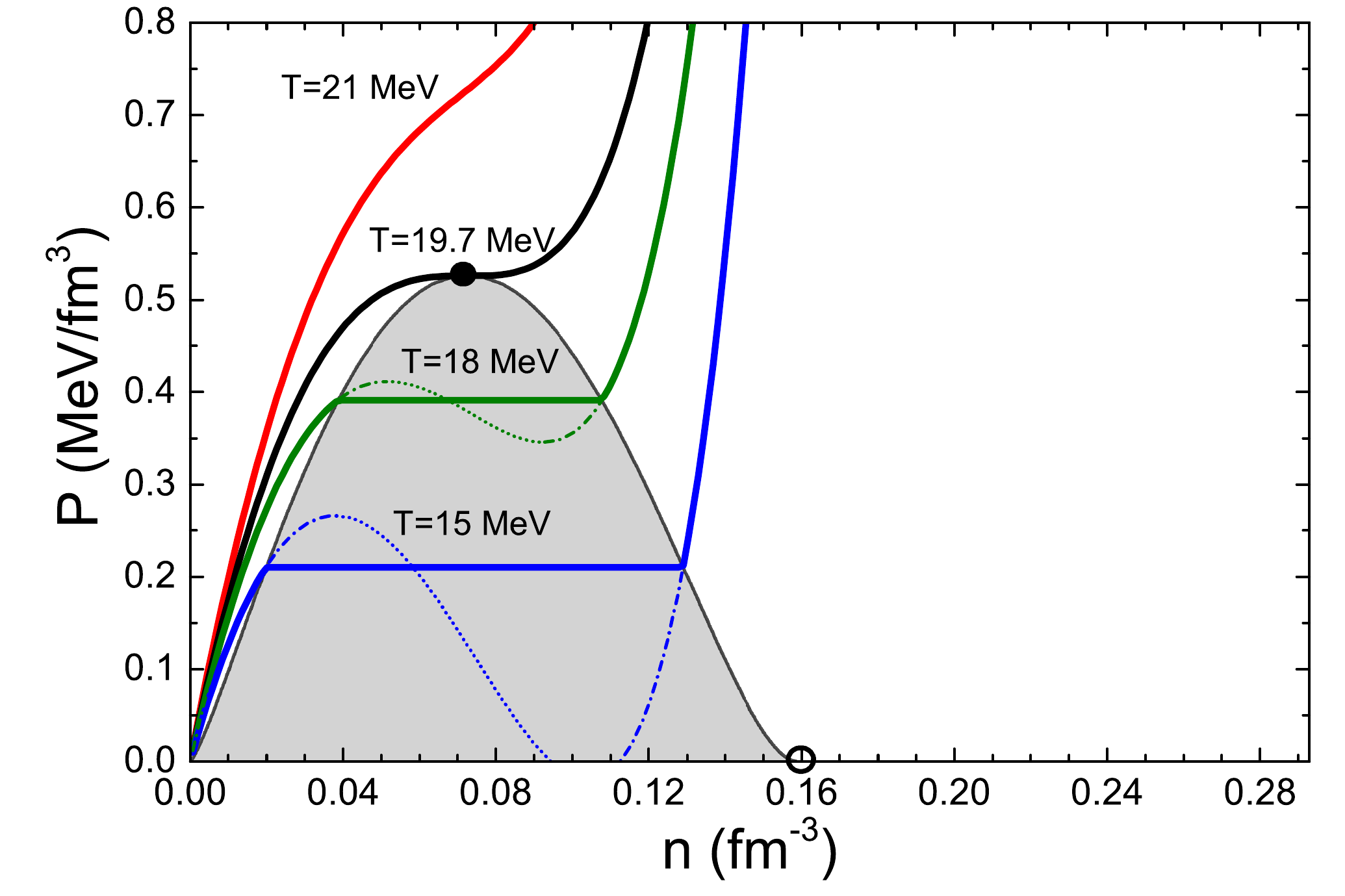}
\includegraphics[width=.49\textwidth]{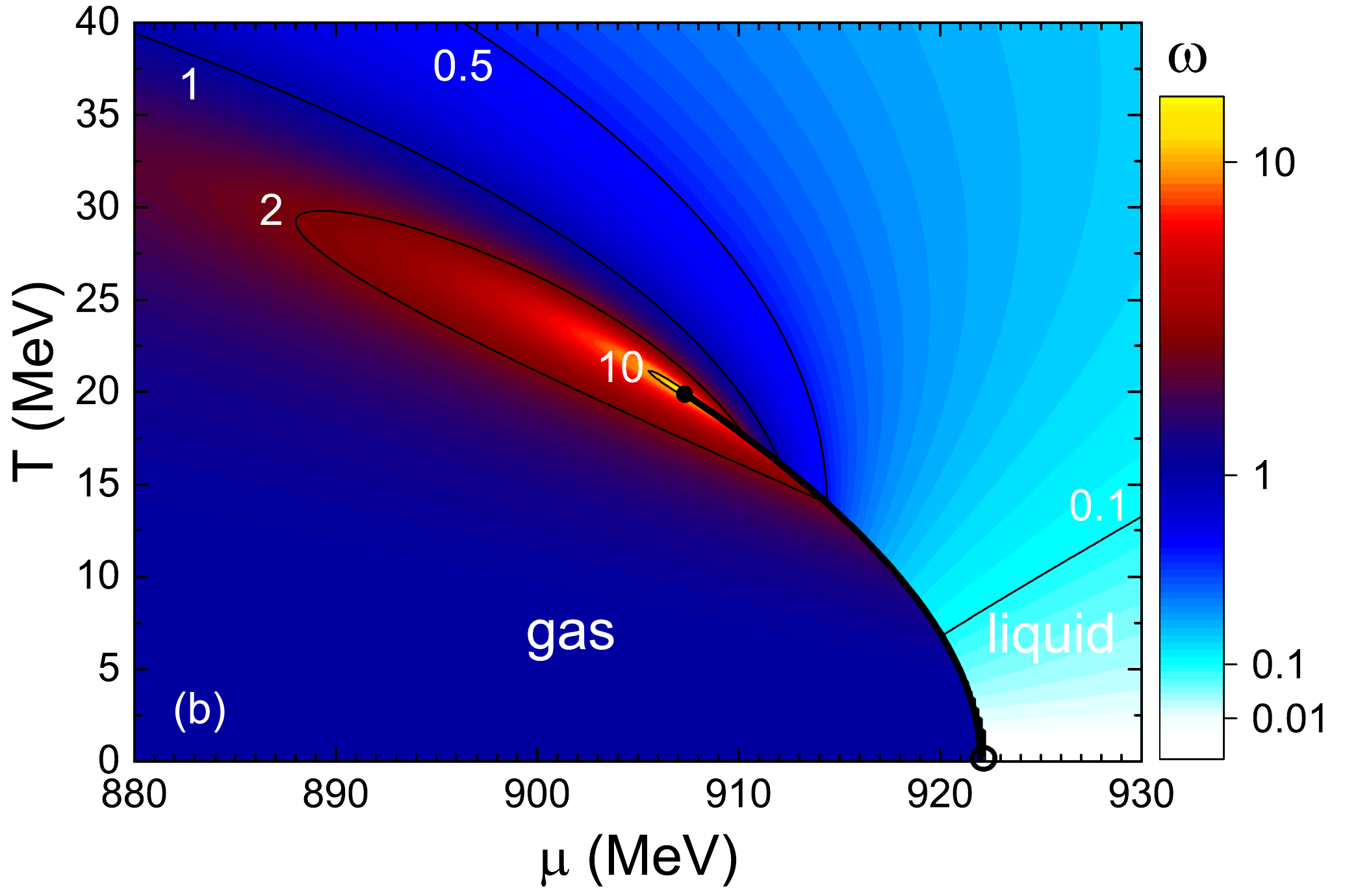}
\caption{
Properties of the symmetric nuclear matter described within the quantum van der Waals equation.
\emph{Left panel:} From Ref.~\cite{Vovchenko:2015vxa}. Pressure-density isotherms. Dash-dotted and dotted lines correspond to metastable and unstable regions, respectively.
\emph{Right panel:} From Ref.~\cite{Vovchenko:2015pya}. Phase diagram in $\mu$-$T$ coordinates. The contour map depicts the scaled variance $\omega$ of nucleon number fluctuations.
The solid and open circles depict the critical point and the nuclear ground state, respectively.
}
\label{fig:NM}
\end{figure}

The QvdW equation can be used to describe basic properties of interacting nucleons -- the nuclear matter.
For this task it is sufficient to keep only the nucleons in Eq.~\eqref{eq:vdw:p}.
The nucleon-nucleon interaction parameters $a$ and $b$ are fixed to reproduce the known nuclear ground state properties, the binding energy of $-16$~MeV and normal nuclear density $n_0$ of $0.16$~fm$^{-3}$ in symmetric nuclear matter, yielding
\eq{\label{eq:vdwparam}
a_{NN} = 329~\text{MeV fm}^3, \qquad b_{NN} = 3.42~\text{fm}^3~.
}
The resulting phase diagram of symmetric nuclear matter is depicted in Fig.~\ref{fig:NM}.
The QvdW model predicts nuclear critical point located at $T_c \simeq 19.7$~MeV and $\mu_c \simeq 908$~MeV~($n_c \simeq 0.05$~fm$^{-3}$).
Qualitatively, the description of nuclear matter within the QvdW equation is quite similar to many other mean field models of nuclear matter.
A comparative study between the QvdW and Walecka models was performed in Ref.~\cite{Poberezhnyuk:2017yhx}, where the two models were shown to be quantitatively very similar at moderate densities up to $n_0$.
One difference to other models is the presence of a limiting density $n_{\rm lim} = 1/b$ in the QvdW model. 
This is a distinctive property of the EV repulsion.

The QvdW model of nuclear matter has been used for a number of applications recently, including 
scaled variance, skewness, and kurtosis of nucleon number fluctuations near the critical point~\cite{Vovchenko:2015pya}, 
a systematic study of quantum statistical effects~\cite{Fedotkin:2019bhq},
the non-congruence of the nuclear liquid-gas transition in presence of two conserved charges~\cite{Poberezhnyuk:2018mwt}, 
the shear viscosity of nuclear matter~\cite{Magner:2016kfl},
and the analytic structure of the grand thermodynamic potential~\cite{Savchuk:2019yxl}.

\subsection{HRG model with a critical point}

The multi-component QvdW model is naturally suited to incorporate the critical point of the nuclear liquid-gas phase transition into the HRG model.
The first such extension~(the QvdW-HRG model) has been formulated in Ref.~\cite{Vovchenko:2016rkn}, where the QvdW interaction terms have been added for all baryon-baryon and antibaryon-antibaryon pairs.
The nucleonic QvdW parameters~[Eq.~\eqref{eq:vdwparam}] have been adopted there for all baryons for simplicity.
At low temperatures and large chemical potentials, where excitations of degrees of freedom other than nucleons can be neglected, this model reduces to the QvdW nuclear matter.
At zero chemical potential, the inclusion of the vdW interactions between baryons leads to a qualitatively different behavior of second and higher moments of fluctuations of conserved charges compared to the ideal HRG model.
For many observables the QvdW-HRG model behavior resembled closely the results obtained from lattice QCD simulations.
It has been pointed out that an improved agreement with the lattice data can be achieved by considering different QvdW parameters for strange and non-strange baryons~\cite{Vovchenko:2017zpj}.

\begin{figure}[t]
\centering
\includegraphics[width=.49\textwidth]{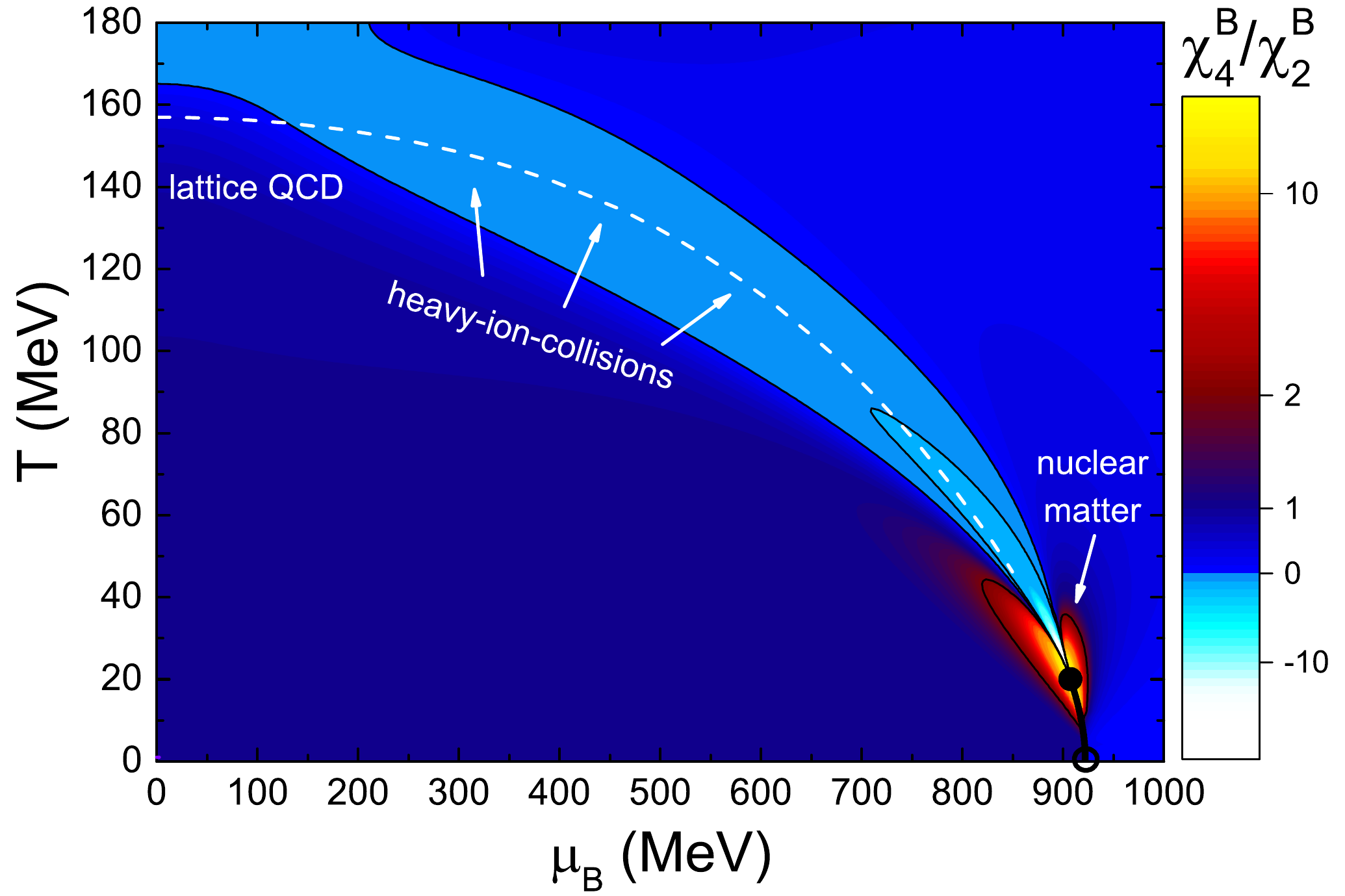}
\includegraphics[width=.49\textwidth]{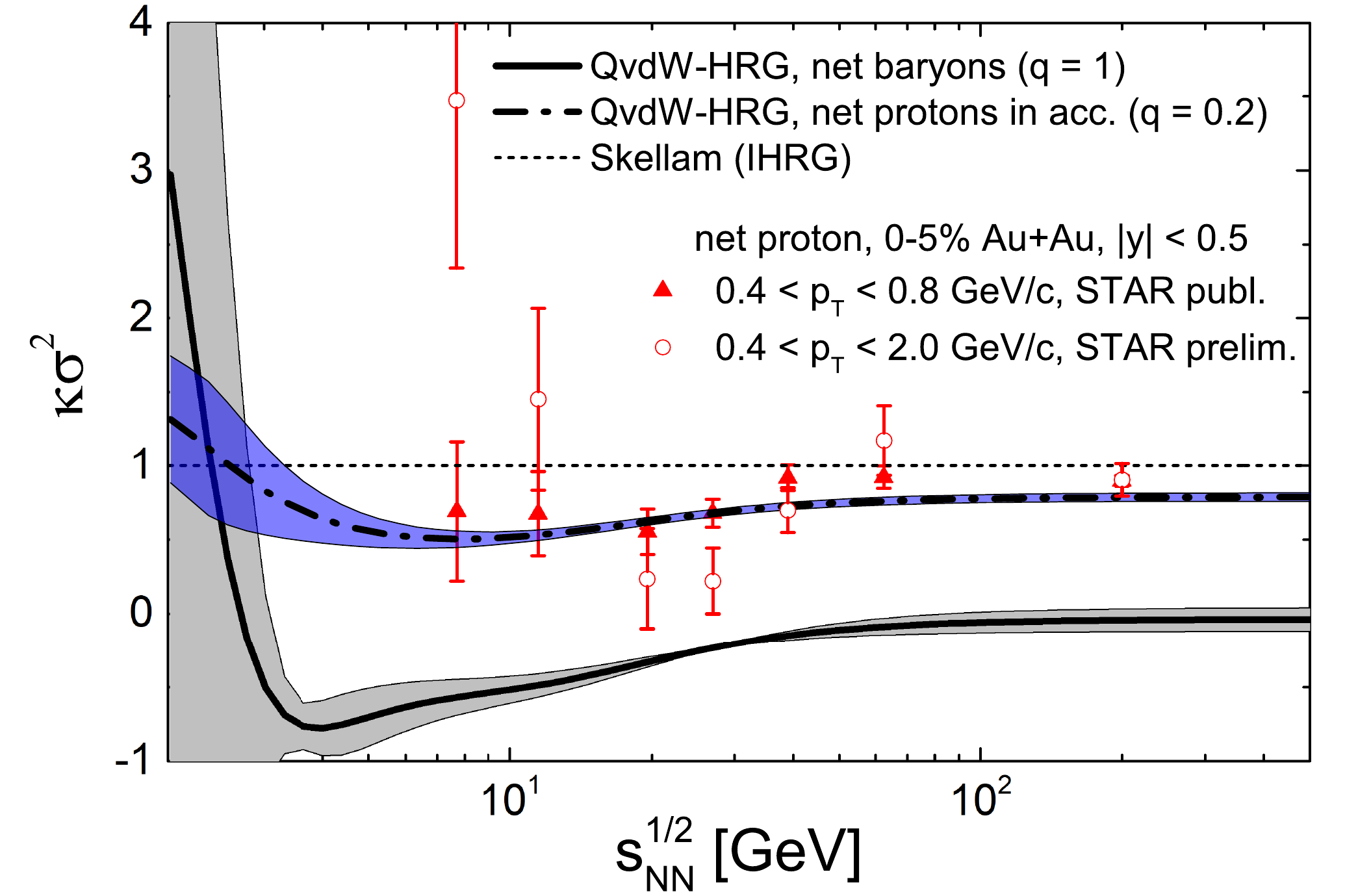}
\caption{
\emph{Left panel:} From Ref.~\cite{Vovchenko:2016rkn}.
Kurtosis of the net-baryon fluctuations within the QvdW-HRG model in the $\mu_B$-$T$ plane.
\emph{Right panel:} From Ref.~\cite{Vovchenko:2017ayq}.
Collision energy dependence of the kurtosis of net baryon and accepted net proton fluctuations, calculated along the phenomenological chemical freeze-out curve~\cite{Cleymans:2005xv}.
Published~\cite{Adamczyk:2013dal} and at that time preliminary~\cite{Luo:2015ewa}~(now finalized~\cite{Adam:2020unf}) data from the STAR collaboration for difference acceptances are shown by full and open red symbols, respectively.
}
\label{fig:vdwhrgkurt}
\end{figure}

The QvdW-HRG model predicted a non-trivial behavior of the higher-order net-baryon and net-proton fluctuations in the regions of the QCD phase diagram probed by heavy-ion collision experiments. 
The behavior stems from the nuclear liquid-gas criticality.
This is illustrated in Fig.~\ref{fig:vdwhrgkurt}, both for the $\mu_B$-$T$ plane, and along the phenomenological chemical freeze-out curve of Ref.~\cite{Cleymans:2005xv}.
Comparison between the kurtosis of net baryons and that of the accepted net protons shown in Fig.~\ref{fig:vdwhrgkurt} reveals significant differences between the two in the presence of the vdW interactions: the net proton kurtosis is considerably closer to the Skellam distribution baseline.
This is different from the ideal HRG model, where both the net proton and net baryon fluctuations are described by the Skellam distribution, and suggests that net proton fluctuations may not necessarily be a good direct proxy for net baryon fluctuations.
An improved and comprehensive analysis has recently been presented in Ref.~\cite{Poberezhnyuk:2019pxs}, where both the hadron yields and the fluctuations of all baryon number, electric charge, and strangeness have been analyzed using the QvdW-HRG model.
The strong effect of the nuclear liquid-gas transition on the fluctuation observables has also been pointed recently in Refs.~\cite{Fukushima:2014lfa,Mukherjee:2016nhb} using relativistic mean-field theory based descriptions.

The QvdW-HRG model has also been considered in the context of mimicking the conjectured QCD critical point rather than the nuclear liquid-gas transition.
In Refs.~\cite{Samanta:2017yhh} and~\cite{Sarkar:2018mbk} the QvdW parameters $a$ and $b$ were fitted to sets of $\mu_B = 0$ lattice data on the equation of state, yielding a critical point location of $T \sim 60-65$~MeV and $\mu_B \sim 700-715$~MeV.
This corresponds to $\mu_B/T \sim 11-12$ at the critical point and is currently beyond the reach of first-principle lattice methods.
It should be noted that the associated phase transition in the QvdW-HRG model is still of a liquid-gas type, which is qualitatively different from the expectations concerning the sought-after chiral QCD critical point~\cite{Steinheimer:2013xxa,Wunderlich:2016aed}.

Other applications and modifications of the QvdW-HRG model include:

\begin{itemize}

\item A possible effect of the vdW interactions on the transport coefficients of hot hadronic matter has been explored in Refs.~\cite{Sarkar:2018mbk,Mohapatra:2019mcl}.

\item A simultaneous inclusion of the vdW interactions and in-medium modifications of hadron masses was worked out in Ref.~\cite{Zhang:2019uct}, suggesting an improved agreement with the lattice data at $T \sim 160-190$~MeV.

\item An influence of the nuclear liquid-gas transition on the analytic properties of the QCD grand potential, in particular on the radius convergence of the Taylor expansion around $\mu_B = 0$, was studied in Ref.~\cite{Savchuk:2019yxl}.
The radius of convergence in the QvdW-HRG model was found to be equal to $r_{\mu_B/T} \sim 2-3$ at temperatures $T\sim140 - 170$~MeV, indicating potentially significant limitations of the Taylor expansion method used in lattice QCD.

\end{itemize}

A more involved way to incorporate the nuclear liquid-gas phase transition into HRG is using relativistic mean-field theory, yielding a relativistically covariant description of an interacting HRG~\cite{Steinert:2018zni}.
Relativistic mean-field theory has also been used as a basis for QCD equation of state descriptions which include both hadronic and partonic degrees of freedom~\cite{Motornenko:2019arp}.

\subsection{Beyond van der Waals}

The vdW equation is perhaps the simplest way to incorporate attractive and repulsive interactions.
More involved mechanisms are required for more involved applications, e.g. for studying the nuclear equation of state at densities exceeding the normal nuclear density.
Over the years, many modifications to the original vdW equation~\eqref{eq:vdwcl} were developed for chemistry-related applications. 
These modifications concern both the attractive and repulsive terms, and yield a class of vdW-like equations of state for real gases.
In Ref.~\cite{Vovchenko:2017cbu} a formalism of quantum statistical real gas models was developed.
The repulsive interactions are treated in terms of a generalized excluded volume whereas the attractive interactions are described by a generalized mean-field.
The free energy of a quantum real gas equation of state reads:
\eq{\label{eq:Frealgas}
F^{\rm rg}(T,V,N) = F^{\rm id} (T, V \, f(n), N) + N \, u(n)~.
}
Here $f(n)$ quantifies the fraction of the total volume which is available for particles to move in at a given value of the particle number density $n$.
$f(n)$ takes values in the range $0 \leq f(n) \leq 1$.
The quantity $u(n)$ is a self-consistent density-dependent mean field, corresponding to intermediate range attractive interactions. All other quantities are calculated from free energy via standard thermodynamic identities (see ~\cite{Vovchenko:2017cbu} for details).
The QvdW model follows as a partial case of Eq.~\eqref{eq:Frealgas} with $f^{\rm vdw} (n) = 1 - bn$ and $u^{\rm vdw}(n) = -an$. 
Many other options are possible, and have been considered in Ref.~\cite{Vovchenko:2017cbu} for the case of symmetric nuclear matter.

\begin{figure}[t]
\centerline{\includegraphics[width=.6\textwidth]{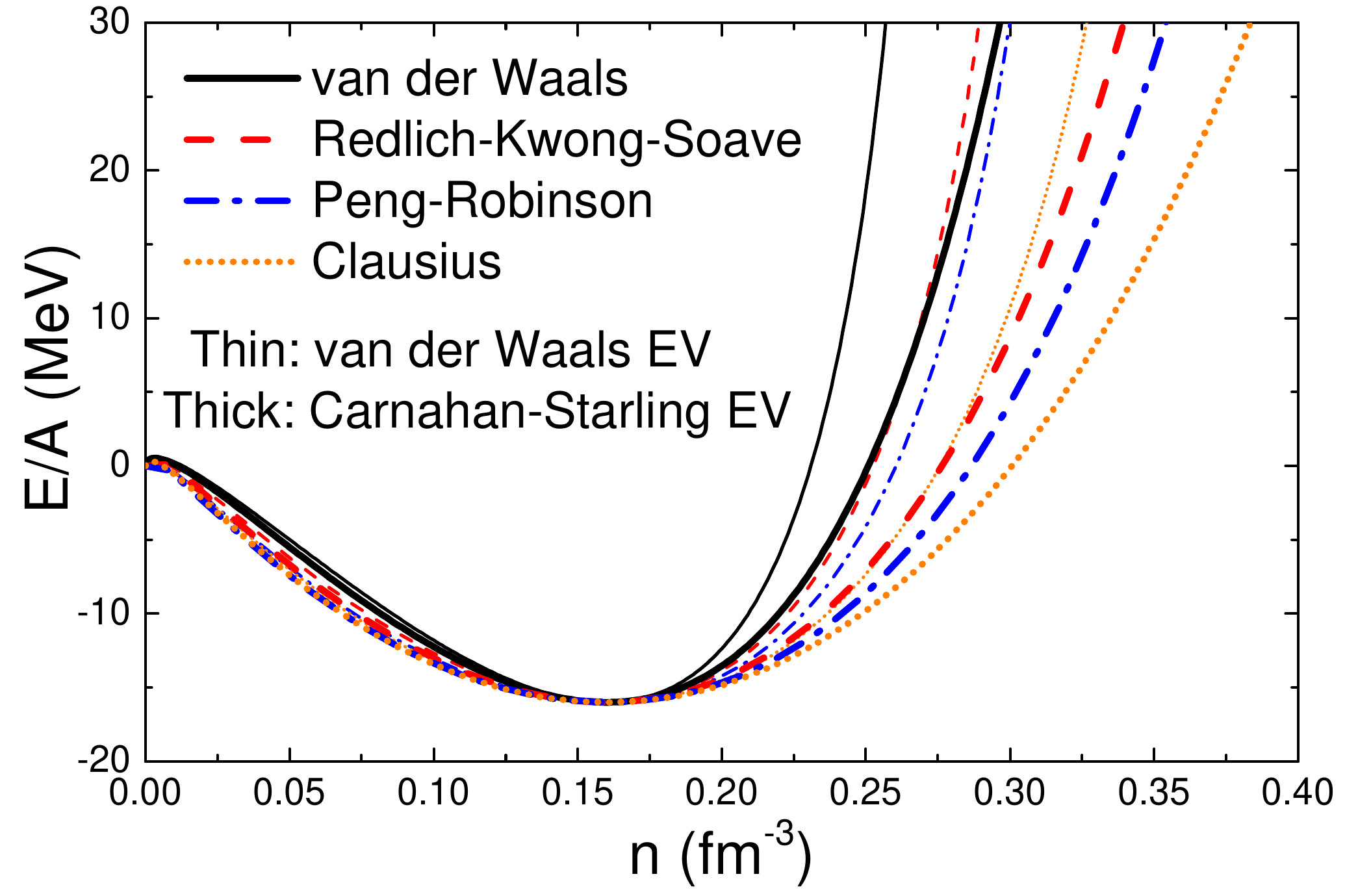}}
\caption{From Ref.~\cite{Vovchenko:2017drx}.
The nucleon number density dependence of the binding energy
per nucleon $E/A$ in symmetric nuclear matter calculated within eight different
real gas models at $T = 0$. The thin lines denote calculations within four models with the vdW EV term, i.e. they correspond to vdW (solid black line),
Redlich-Kwong-Soave (dashed red line), Peng-Robinson (dash-dotted blue line),
and Clausius (dotted orange line) models. The thick lines correspond to models
with the Carnahan-Starling EV term~\cite{carnahan1969equation}.
}
\label{fig:binding}
\end{figure}

Figure~\ref{fig:binding} depicts the density dependence of the nuclear binding energy at $T = 0$ in various quantum statistical real gas models.
Real gas models can be used to improve the description of the nuclear equation of state at densities above $n_0$.
They allow to bring  the nuclear ``incompressibility''
factor $K_0$ down from a high QvdW model value of $762$~MeV closer to empirical estimates.

An interesting interpretation of the real gas model formalism was given in Ref.~\cite{Lourenco:2019ist}, where the real gas equations of state are regarded as a density-dependent vdW equation of state.
Indeed, the free energy of a real gas~\eqref{eq:Frealgas} can be cast in the vdW form~\eqref{eq:Fvdw} with density-dependent vdW parameters:
\eq{
a(n) = -\frac{u(n)}{n}, \qquad b(n) = \frac{1 - f(n)}{n}~.
}
Ref.~\cite{Lourenco:2019ist} considered a density-dependent vdW~(DD-vdW) model, comprising of a Carnahan-Starling treatment of the repulsive interactions~\cite{carnahan1969equation} and a Clausius model inspired attractive mean-field~\cite{Vovchenko:2017ygz}.
The resulting DD-vdW equation of state was shown to satisfy many nuclear and neutron matter constraints, including the recent data from the GW170817 neutron star merger event~\cite{TheLIGOScientific:2017qsa,Abbott:2018exr}.
Another vdW model approach constrained to a number of nuclear matter constraints makes use of an induced surface tension concept~\cite{Ivanytskyi:2017pkt}.

\section{Summary and outlook}

Much progress has been made in recent years in understanding the equation of state of hadronic phase and the development of the hadron resonance gas model.
The onset of deviations from an uncorrelated gas of hadrons behavior seen in various observables in lattice QCD can be understood in terms of excluded volume like, repulsive interactions between baryons.
Future lattice data on susceptibilities of conserved charges and imaginary-$\mu$ Fourier coefficients is suited to provide further, quantitative constraints on repulsive hadronic interactions.
This will, in turn, clarify their relevance for the heavy-ion observables like yields and fluctuations of identified hadrons.

The recently developed quantum van der Waals theory of nuclear and hadronic interactions allowed to elaborate in some detail the role of the nuclear liquid-gas transition and the associated critical point in the properties of hot QCD, the conserved charges susceptibilities in particular.
A significant role of the nuclear matter CP in net-proton and net-charge fluctuations measured in low and intermediate energy heavy-ion collisions seems inescapable.
More quantitative statements require an implementation of critical dynamics in simulations of heavy-ion collisions.

The vdW model has also proven as a quite useful tool to study various phenomena associated with the presence of a critical point, which is particularly interesting in the context of the ongoing search for the QCD critical point.
One potentially interesting and so far unexplored possibility is the vdW model in a finite volume, where one could study how the critical phenomena reflect themselves in finite systems relevant for heavy-ion collision experiments.
Another potential avenue is molecular dynamics simulations of vdW fluids, which can shed light on non-equilibrium effects.

The various excluded-volume and van der Waals generalizations of the hadron resonance gas model presented in this review are implemented within a publicly available \texttt{Thermal-FIST} package~\cite{Vovchenko:2019pjl}.

\section*{Acknowledgments}

The author acknowledges helpful comments from  B.~D\"onigus, M.I.~Gorenstein, and V.~Koch.
The author also thanks P.~Alba, D.~Anchishkin, A.~Motornenko, L.~Jiang, R.~Poberezhnyuk, L.M.~Satarov, and H.~Stoecker for many useful discussions on the hadron resonance gas model and collaboration on related projects.

\bibliographystyle{ws-ijmpe}
\bibliography{QvdW-review}

\end{document}